# Spectroscopic Evidence for Charge Order Melting via Quantum Fluctuations in a Cuprate

W. S. Lee[1]*, K. J. Zhou[2*], M. Hepting[1], J. Li[2], A. Nag[2], A. C. Walters[2], M. Garcia-Fernandez[2], H. Robarts[2], M. Hashimoto[3], H. Lu[4], B. Nosarzewski[4], D. Song[5], H. Eisaki[5], Z. X. Shen[1,6], B. Moritz[1], J. Zaanen[6], T. P. Devereaux[1]*

[1]Stanford Institute for Materials and Energy Sciences, SLAC National Accelerator Laboratory and Stanford University, 2575 Sand Hill Road, Menlo Park, California 94025, United States of America

[2]Diamond Light Source, Harwell Science and Innovation Campus, Didcot, Oxfordshire OX11 0DE, United Kingdom

[3]Stanford Synchrotron Radiation Lightsource, SLAC National Accelerator Laboratory, Menlo Park, CA 94025, United States of America

[4]Department of Physics, Stanford University, Stanford, California 94305, United States of America

[5]National Institute of Advanced Industrial Science and Technology (AIST), Tsukuba, Ibaraki 305-8560, Japan

[6]Geballe Laboratory for Advanced Materials, Departments of Physics and Applied Physics, Stanford University, Stanford, CA 94305, USA.

[7]Instituut-Lorentz for theoretical Physics, Leiden University, Niels Bohrweg 2, 2333 CA Leiden, The Netherlands.

Correspondence to: leews@stanford.edu, kejin.zhou@diamond.ac.uk, tpd@stanford.edu

**Copper-oxide high $T_C$ superconductors possess a number of exotic orders co-existing with or proximal to superconductivity, whose quantum fluctuations may account for the unusual behaviors of the normal state, even affecting superconductivity [1-4]. Yet, spectroscopic evidence about such quantum fluctuations remains elusive. Here, we reveal spectroscopic fingerprints for such fluctuations associated with a charge order (CO) [5-14] in nearly optimally-doped $Bi_2Sr_2CaCu_2O_{8+\delta}$, using resonant inelastic x-ray scattering (RIXS). In the superconducting state, while the quasi-elastic CO signal decreases with temperature, the interplay between CO fluctuations and bond-stretching phonons in the form of a Fano-like interference paradoxically increases, incompatible with expectations for competing orders. Invoking general principles, we argue that this behavior reflects the properties of a dissipative system near an order-disorder quantum critical point, where the dissipation varies with the opening of the pseudogap and superconducting gap at low temperatures, leading to the proliferation of quantum critical fluctuations which melt CO.**

Charge order (CO), which is ubiquitous in hole-doped cuprates [5-14], is accompanied by a negligible lattice deformation (~ 0.1 pm [15]); however, signatures of valence electron density modulations due to CO can be detected by RIXS at the Cu *L*-edge. RIXS resolves both the quasi-static and dynamical inelastic signals [8,16], as highlighted in Fig. 1a, particularly the intensity of the inelastic branch of excitations below 0.1 eV. These excitations possess a similar energy scale as bond-stretching phonons, which exhibit anomalous softening and broadening in certain portions of reciprocal space, observed using inelastic neutron scattering (INS) and non-resonant inelastic x-ray scattering (IXS) [18, 19]. These behaviors have suggested a coupling with CO [18,19] and possibly some form of charge collective mode [20]. However, while INS and IXS measure the phonon self-energy (*i.e.* dynamical structure factor), RIXS largely reflects the electron-phonon coupling itself and interference with charge excitations [16,17]. With superb momentum resolution, RIXS at the Cu *L*-edge already has revealed two distinct anomalies associated with CO excitations due to a Fano-like interference effect [17] as shown in Fig. 1b: (i) an apparent softening of the RIXS phonon at the CO wave-vector ($Q_{CO}$), and (ii) creation of a "funnel"-like spectral weight emanating from $Q_{CO}$ with a non-monotonic integrated momentum-distribution whose maximum occurs at $Q > Q_{co}$.

Figure 1c shows the temperature evolution of CO in the quasi-elastic region, (Methods and Supplementary Fig. 2) with corresponding energy integrated CO peak profiles in Fig. 1d. A broad peak centered at ~ 0.25 reciprocal lattice units (*r.l.u*) already exists at 260 K, essentially T* the psesudogap temperature [21]. As the temperature decreases, the CO peak height gradually increases and the width narrows (Fig. 1e), consistent with an increase in both the amplitude and correlation length of the order parameter. The CO peak height reaches a maximum at $T_C$ and then decreases precipitously upon entering the superconducting state (becoming less than 1/3 of the value at 90 K), but with a correlation length of ~ 30 Å at 15 K. These observations are consistent with temperature evolution of CO in $YBa_2Cu_3O_{6+x}$ [8,9] and near-optimally doped $La_{1-x}Sr_xCuO_4$ [14,15], where reduction of CO for $T < T_C$ was interpreted in terms of order parameter competition with superconductivity.

A very different story emerges from the higher energy inelastic response as shown in the raw data of Fig. 2a. Upon cooling from 260 K, the RIXS phonon dispersion gradually softens roughly centered around $Q_{CO}$, from ~ 0.2 to 0.35 *r.l.u.* (Fig. 2b). Below $T_C$, this softening becomes more pronounced and a "dip" in the dispersion develops at $Q_{CO}$ (Fig. 2b), in contrast to the significant reduction in quasi-elastic CO (Fig. 1 and Fig. 2a). Between 260 K and 15 K, the RIXS phonon peak energy at ~ $Q_{CO}$ shifts from ~ 58 meV to 36 meV (inset of Fig. 2b). Such shift is evident even in the raw data without applying any fitting (Fig. 2c). Both the funnel-like spectral weight (Fig. 2d) and integrated momentum distribution curve (MDC) for the phonon intensity (Fig. 2e) exhibit significant enhancement once the system enters the superconducting state.

These behaviors are different from the phonon self-energy evolution in the superconducting state probed by IXS and INS. The opening of a superconducting gap in the particle-hole continuum (here $2\Delta \sim 80$ meV) should *sharpen* the bond-stretching phonon spectral width [21], as its energy scale lies below $2\Delta$; however, the RIXS phonon width near $Q_{CO}$ *broadens,* with no obvious changes discernible elsewhere (Fig. 2f). Neither can they be explained by a Kohn anomaly [22], which would require unrealistically large electron-phonon coupling strength to produce the observed magnitude of softening. In addition, the Kohn anomaly should decrease concomitant to the reduction in CO magnitude, as seen in the low energy phonon anomalies in YBCO via IXS [24]. These again testify to the fact that the RIXS phonon cross-section is not reflective of the phonon self-energy, but rather the electron-phonon coupling itself and its interference with underlying charge excitations [16, 17]. The enhancement of the RIXS phonon anomalies near $Q_{CO}$ indicates that the CO excitations' spectral weight increases in the superconducting state. Yet, the contrasting temperature evolution shown in Figs. 1 and 2 reveals a major paradox, which departs from the picture of order parameter competition as in Landau theory where the reduction of CO due to superconductivity ultimately would lead to a *weakening* of the CO excitations' spectral weight, in contrast with our observations.

We propose a natural way to understand this temperature dependence paradox. Due to a much smaller mass for electrons than that of atoms forming solids, an electronic crystal, such as the CO, can be subjected to severe quantum zero-point motion that can reduce the magnitude of the order. The fact that the weight of the CO excitations at the elevated energies of the bond-stretching phonon decouples from the magnitude of the static order parameter may signal that the system lies in close proximity to a strong coupling, order-disorder, "phase"-driven quantum phase transition, rather than one driven by amplitude fluctuations [25]. This is analogous to "dynamical stripes", which were based on a similar organization of the spin excitations measured by inelastic neutron scattering [26-27] in the La-family of cuprate superconductors. As shown in Fig. 3 at $g_2$, which is far away from the quantum critical point (QCP at $g_C$) on the ordered side (denoted renormalized classical, R.C.) of the phase transition, excitations can be characterized by (i) a static order parameter, (ii) the associated Goldstone bosons, and (iii) a "branch-cut" continuum associated with quantum fluctuations of the quantum critical state (red shading in Fig. 3), starting at ~ $kT_O$ and extending to high energy ($T_O$ denotes the transition temperature in the R.C. region). At the QCP, both the static order and Goldstone bosons vanish, with spectral weight transferred into the continuum.

In practice, this should be a dissipative quantum phase transition [28] because of the metallic-like particle-hole excitations that can suppress quantum fluctuations. Here, the dissipation should be strongly temperature dependent: the opening of the pseudogap below T* and a clean gap in the superconducting state, significantly reduces dissipation at low temperatures. We are unaware of explicit calculations dealing with this motif, as one usually considers a fixed dissipation parameter at zero temperature. However, it is easy to see the qualitative outcome (Fig. 3). At a given temperature, one can "freeze" the dissipation parameter α and compute the "distance" from the QCP, $g_c(\alpha)$. Keeping the coupling constant fixed at $g_1$ and lowering the temperature, or in effect decreasing the dissipation parameter, the system effectively moves toward the QCP. In the context of our experimental observations, entering the superconducting state suppresses dissipation such that quantum fluctuations become more severe and "melt" the CO.

We can model the resulting RIXS spectrum as a superposition between a discrete dispersive mode (bond-stretching phonon) and a continuum of quantum fluctuations emanating from $Q_{CO}$. Well known from Raman scattering, any coupling between a discrete mode and a continuum will lead to a distortion of the combined spectra, resulting in a Fano-like lineshape. This occurs where the phonon branch intersects the continuum near $Q_{CO}$. The spectra can be modeled remarkably well using this simple and natural construction, where we take linearly dispersing electronic modes with fixed spectral weight, attenuated by a temperature dependent damping constant. Their propagator takes the form of an anti-Lorentzian

$$\chi(q,\omega) = \frac{1}{\omega - \Omega_q + i\Gamma(T)} - \frac{1}{\omega + \Omega_q + i\Gamma(T)}$$

where $\Omega_q = c \cdot |q - Q_{CO}|$ is the CO excitation energy, c is the mode velocity, and $\Gamma(T)$ is the damping, similar to the standard form for quantum critical excitations. To account for the effect of the finite CO correlation length on the continuum, we apply a Gaussian momentum broadening on $\chi$, whose FWHM corresponds to the CO width shown in Fig. 1e (Methods). The Fano-effect explains the CO-induced anomalies in the RIXS phonon cross-section (*e.g.* Fig. 4a and b), where the coupling reorganizes spectral weight according to the strength of the continuum. Note that the apparent "Fano phonon" (*i.e.* RIXS phonon) dispersion can significantly deviate from the actual phonon dispersion due to only the phonon self-energy in the momentum region where the Fano-interference is strong. By adjusting only the temperature-dependent damping constant $\Gamma(T)$ (Methods and Supplementary Fig. 3), we can achieve excellent agreement with our temperature dependent data, including the energy-momentum intensity map (Fig. 4c), phonon dispersion (Fig. 4d), and the integrated phonon intensity MDC (Fig. 4e). From the fit, as the temperature decreases, the damping decreases, and appears to do so more rapidly when approaching $T_C$ (Fig. 4f), presumably due to the opening of the superconductivity gap

Interestingly, the increment of Γ as a function of temperature is of the same order of magnitude as the "Planckian" dissipation [4], $kT/\hbar$, consistent with our conjecture of dealing with strongly interacting quantum criticality [22]. We note that the zero-temperature residue $\Gamma_o \sim 40$ meV may be related to a disorder pinning gap or the effective gap in the quantum critical spectra, which are beyond the resolution of our current data. Higher resolution RIXS spectra would allow us to gain further insight into the lower energy structure of the electronic continuum.

Our results provide strong support for the existence of quantum phase fluctuations inside the superconducting dome, with the corresponding CO QCP located near the optimal doping in our Bi2212 sample. Furthermore, our proposed dissipation-driven quantum melting of CO provides a new perspective on the relationship between CO and SC, which is beyond a simple competition. It may provide a natural explanation for the lack of CO reduction inside the superconducting state of some other cuprates. Experiments on single-layer Bi-based cuprates (Bi-2201, Ref. [6]) and underdoped $La_{2-x}Sr_xCuO_4$ [14] tend to show CO largely insensitive to superconductivity. This lack of quantum melting of CO implies that proximity to the CO QCP may be material dependent; and for some materials, the QCP even may lie beyond the superconducting dome [29,30]. Finally, a broad CO feature still exists at the pseudogap temperature T*, and the excitations already plays a role at high temperatures, where one finds pseudogap and the strange metal behavior [1]. So, it may well be that the pseudogap and the strange metal behaviors are intimately connected with this quantum critical CO state.

**Methods:**

**<u>Materials</u>**

High-quality Bi-2212 ($Bi_{1.7}Pb_{0.4}Sr_{1.7}CaCu_2O_{8+\delta}$) single-crystals were grown by floating-zone (FZ) methods. The crystals were annealed in nitrogen atmosphere to achieve a nearly-optimally doped hole concentration with Tc = 90 K; this corresponds to a doping concentration of ~ 13 % holes per Cu cation. Flat crystals with a sharp Laue pattern were selected for the reported RIXS experiments.

**Ultrahigh high resolution RIXS measurements**

The RIXS measurements were performed using the RIXS spectrometer at the ID21-RIXS beamline of the Diamond Light Source in the United Kingdom. The RIXS spectra were taken with the photon energy of the incident x-rays tuned to the maximum of the absorption curve near the Cu $L_3$-edge (Supplementary Fig. 1a). The scattering geometry is sketched in Supplementary Fig. 1b. The data were collected with a linear vertical polarization (σ-polarization) of the incident beam. The energy resolution was approximately $\Delta E$ ~ 35 meV (FWHM) at Cu $L_3$-edge. The scattering angle of the spectrometer was set at $2\theta = 146^{\circ}$. Since the electronic state in Bi2212 is quasi-two-dimensional, *i.e.* rather independent along the *c*-axis, the data shown in this report are plotted as a function of in-plane projected momentum transfer $q_{//}$, *i.e.* projection of $q = k_f - k_i$ onto the $CuO_2$ plane. Different in-plane momentum transfers, $q_{//}$ (projection of the scattering vector $q$ along [100]), were obtained by rotating the samples around the vertical *b*-axis. Note that the scattering vector $q$ is denoted using the pseudotetragonal unit cell with $a = b = 3.82$ Å and $c = 30.84$ Å, where the *c*-axis is normal to the sample surface. In our convention, positive $q_{//}$ corresponds to grazing-emission geometry and negative $q_{//}$ corresponds to grazing-incidence geometry. For all RIXS data the $q_{//}$ resolution is smaller than 0.02 r.l.u.

**Data analysis and fitting**

The data were normalized to incident photon flux and corrected for self-absorption effects using the formalism described in the Supplemental Material of Ref. 31. The zero energy positions were determined by comparing the spectrum recorded from a small amount of silver paint (at each $q_{//}$) near the sample surface, and fine adjusted by the fitted elastic peak position, as shown in the Supplementary Fig 2a. The fitting model involves a Gaussian (elastic peak), an anti-Lorentzian (phonon peak) and a background using the tail of an anti-symmetrized Lorentzian. The model is convolved with the energy resolution of the RIXS instrument (FWHM = 35 meV, Gaussian convolution) and fit to the data. The results of these fits are presented in Supplementary Fig. 2b.

From this fitting analysis, we extracted the dispersion (peak positions) of the RIXS phonon shown in Fig. 2b. In addition, we also obtained the quasi-elastic maps in Fig. 1b, c and phonon maps in Fig. 2d by respectively subtracting the fitted phonon and elastic peaks from the raw data.

The CO peak profiles shown in Fig. 1e were fitted to a Gaussian peak with a linear background. For T = 130 K and 220 K data, the linear background is estimated by linear interpolation between the 90 K and 260 K data. The error bars of 130K and 220 K data were estimated by the difference of fitted parameter value if use T = 90 K and 260 K background, respectively.

**Model simulation of Fano effect between CO excitations and the bond stretching phonons**

An explicit calculation for the RIXS spectral function involves a 4-particle correlation function, with an accurate description of the quantum critical charge ordered state for the cuprates. Since such a description is not currently available, we instead appeal to general phenomenology and model the RIXS spectra as having a Fano lineshape whereby the phonon seen in RIXS interacts with a charge particle continuum, *i.e.* CO collective modes. The Fano spectra were modeled using the form of the Raman response [32] with a straightforward generalization to finite momentum $q$:

$$\chi''_{Fano}(q,\omega) = \frac{(\omega+\omega_a)^2}{(\omega^2-\tilde{\omega}_\lambda^2)^2+[2\omega_\lambda\Gamma_\lambda(\omega)]^2}\left\{\gamma_\lambda^2\chi_\lambda''\left[(\omega-\omega_a)^2+4\Gamma_\lambda^i\Gamma_\lambda(\omega)\left(\frac{\omega_\lambda}{(\omega+\omega_a)}\right)^2\right]+\right.$$

$$\left.4g_{pp}^2\Gamma_\lambda^i\left(\frac{\omega_\lambda}{(\omega+\omega_a)}\right)^2\left[1+\frac{\lambda(\omega)}{\beta}\right]^2\right\}.$$

$\omega_a = \omega_\lambda\sqrt{1+\beta}$ sets the position of the antiresonance of the lineshape with $\beta = \frac{2g_{pp}g_\lambda}{\gamma_\lambda\omega_\lambda}$, where $g_\lambda$ is electron-phonon coupling vertex, $\omega_\lambda$ is the bare phonon frequency and $\gamma_\lambda$ is a symmetry element of the Raman vertex projected out by the incoming and outgoing photon polarization vectors. $\chi_\lambda(\omega)$is the complex electronic susceptibility representing the CO-excitations that causes the Fano effect on the phonon spectrum. The renormalized phonon frequency is defined through the following equation $\tilde{\omega}_\lambda^2 = \omega_\lambda^2(1-\lambda(\omega))$, where $\lambda(\omega) = 2g_\lambda^2\,\chi_\lambda'/\omega_\lambda$ and $\chi_\lambda'$denotes the real part of $\chi_\lambda$. $\Gamma_\lambda$ is renormalized line width of phonon, $\Gamma_\lambda = \Gamma_\lambda^i + g_\lambda^2\chi_\lambda''$ where $\chi_\lambda''$ represent the imaginary part of $\chi_\lambda$. And $\Gamma_\lambda^i$is the intrinsic damping of the phonon away from $Q_{CO}$. Finally, $g_{pp}$ represent the photon-phonon coupling which in principle is derivable from the direct resonant matrix elements [17].

Although there are many parameters involved in this fitting procedure, many of them can be well-constrained by the data. For example, the renomalized phonon frequency and linewidth can be obtained from the energy and momentum position of the phonon intensity near $Q_{CO}$, while the bare phonon parameters can be obtained away from $Q_{CO}$. The electronic response $\chi_\lambda$ is determined by the response at low energies near $Q_{CO}$, given the parameters of the CO collective modes. Finally, $g_{pp}$ sets the overall intensity of the phonon relative to the continuum and $\beta$ sets the position of the anti-resonance (depletion of spectral weight) on the high energy side of the phonon.

In this calculation, the momentum dependent electron-bond-stretching phonon coupling takes the expected form $g_\lambda = g_{gg}sin\left(\frac{\pi q}{2}\right)$ for deformational coupling. Constrained by the bond-stretching phonon energies at small q and near the zone boundary, we model the phonon dispersion $\omega_\lambda(q) = 0.08 - 0.014 sin\left(\frac{\pi q}{2}\right)$ in units of electron volts (eV). $g_{pp}$ is expected to have the same functional form as $g_\lambda$, thus, $g_{pp} = g_{pp} sin\left(\frac{\pi q}{2}\right)$. Taking the simplest form for the electronic susceptibility of CO excitations, we assume $\chi(q,\omega) = \frac{1}{\omega - \Omega_q + i\Gamma(T)} - \frac{1}{\omega + \Omega_q + i\Gamma(T)}$, where $\Omega_q = c \cdot |(q - Q_{CO})|$, c, and $\Gamma(T)$ denote the energy of the mode, the velocity, and the damping, respectively. We use c = $\sqrt{0.4}$ eVπ with good agreement with the data. In addition, to account for the finite correlation length of CO, we also apply a Gaussian broadening on $\chi$ with FWHM corresponding to the CO width (Fig. 1e) to obtain the continuum of the CO quantum fluctuations (Fig. 1b). To determine the damping Γ, we fit the peak positions of the Fano spectrum near $Q_{CO}$ to those in the data shown in Fig. 2b, while fixing other parameters.

**References:**

1. Keimer, B. *et al.,* From quantum matter to high-temperature superconductivity in copper, *Nature* **518,** 179–186 (2015).
2. Varma, C. M. Pseudogap phase and the quantum-critical point in copper-oxide metals, *Phys. Rev. Lett.* **83**, 3538 (1999).
3. Sachdev, S. Where is the quantum critical point in the cuprate superconductors, *Physica Status Solidi B* **247**, 537 (2010).
4. Zaanen, J. Why the temperature is high. *Nature* **430**, 512 (2004).
5. Tranquada, J. M., Sternlieb, B. J., Axe, J. D., Nakamura, Y. & Uchida, S. Evidence for stripe correlations of spins and holes in copper oxide superconductors. *Nature* **375**, 561 – 563 (1995).

6. Howald, C., Eisaki, H., Kaneko, N. & Kapitulnik, Coexistence of periodic modulation of quasiparticle states and superconductivity in $Bi_2Sr_2CaCu_2O_{8+\delta}$. *Proc. Natl Acad. Sci. USA* **100**, 9705 – 9709 (2003).

7. Abbamonte, P. *et al.,* Spatially modulated 'Mottness' in $La_{2-x}Ba_xCuO_4$. *Nat. Phys.* **1**, 155 – 158 (2005).

8. Ghiringhelli, G. *et al.,* Long-range incommensurate charge fluctuations in $(Y,Nd)Ba_2Cu_3O_{6+x}$. *Science* **337**, 821 – 825 (2012).

9. Chang, J. *et al.,* Direct observation of competition between superconductivity and charge density wave order in $YBa_2Cu_3O_{6.67}$. *Nat. Phys.* **8**, 871 – 876 (2012).

10. Comin, R. *et al.,* Charge order driven by fermi-arc instability in $Bi_2Sr_{2-x}La_xCuO_{6+\delta}$. *Science* **343**, 390 – 392 (2014).

11. da Silva Neto, E. H. *et al.,* Ubiquitous interplay between charge ordering and high-temperature superconductivity in cuprates. *Science* **343**, 393 – 396 (2014).

12. Tabis, W. *et al.,* Charge order and its connection with Fermi-liquid charge transport in a pristine high-$T_c$ cuprate. *Nat. Commun.* **5**, 5875 (2014).

13. Croft, T. P. *et al.,* Charge density wave fluctuations in $La_{2-x}Sr_xCuO_4$ and their competition with superconductivity. *Phys. Rev.* B **89**, 224513 (2014).

14. Wen, J.-J. *et al*., Observation of two types of charge density wave orders in superconducting $La_{2-x}Sr_xCuO_4$, *Nat. Commun.* **10***,* 3269 (2019).

15. Forgan, F.M. *et al.* The microscopic structure of charge density wave in underdoped $YBa_2Cu_3O_{6.54}$ revealed by x-ray diffraction, *Nat. Commun.* **6**, 10064 (2015).

16. Chaix, L. *et al*., Dispersive charge density wave excitations in $Bi_2Sr_2CaCu_2O_{8+d}$, *Nat. Phys.* **13**, 952-956 (2017).

17. Devereaux, T. P. *et al.*, Directly characterizing the relative strength and momentum dependence of electron-phonon coupling using resonant inelastic x-ray scattering. *Phys. Rev. X* **6**, 041019 (2016).

18. Reznik, D. Phonon anomalies and dynamic stripes. *Physica C* **481**, 75 (2012).

19. Pintschovius L. & Braden, M. Anomalous dispersion of LO phonons in $La_{1.85}Sr_{0.15}CuO_4$. *Phys. Rev. B* **60**, R15039 (1999).

20. Park, S. R. *et al.*, Evidence for a charge collective mode associated with superconductivity in copper oxides from neutron and x-ray scattering measurements of $La_{2-x}Sr_xCuO_4$. *Phy. Rev. B* **89**, 020506 (R) (2014).

21. Hashimoto, M., Vishik, I. M., He, R. H., Devereaux, T. P., & Shen, Z. X. Energy gaps in high-transition-temperature cuprate superconductors. *Nat. Phys.* **10**, 483 – 495 (2014).

22. Kohn, W. Image of the Fermi surface in the vibrational spectrum of a metal. *Phys. Rev. Lett.* **2**, 393-394 (1959).

23. Kim, H. H. *et al.* Uniaxial pressure control of competing orders in a high-temperature superconductor, *Science* **362**, 1040-1044 (2014).

24. Allen, P. B. *et al.*, Neutron-scattering profile of Q ≠ 0 phonons in BCS superconductors. *Phys. Rev. B* **56**, 5552 (1997).

25. Sachdev, S. *Quantum Phase transition*, Cambridge University Press, Cambridge, UK (2011).

26. Vojta, M. Lattice symmetry breaking in cuprate superconductors: stripes, nematics, and superconductivity. *Adv. Phys.* **58**, 699-820 (2009).

27. Kivelson, S. A. *et al.,* How to detect fluctuating stripes in the high-temperature superconductors. *Rev. Mod. Phys.* **75**, 1201–1241 (2003).

*28.* Schmid, A. Diffusion and Localization in a Dissipative Quantum System. *Phy. Rev. Lett.,* **51,** 1506 (1983).

29. Peng, Y. Y. *et al.* Re-entrant charge order in overdoped $(Bi,Pb)_{2.12}Sr_{1.88}CuO_{6+\delta}$ outside the pseudogap regime, *Nat. Matter.* **17**, 607 (2018).

30. Webb, T. A. *et al.* Density wave probes cuprate quantum phase transition, *Phys. Rev. X* **9**, 021021 (2019).

**References Cited in Methods:**

31. Minola, M. *et al.,* Collective nature of spin excitations in superconducting cuprates probed by resonant inelastic x-ray scattering. *Phys. Rev. Lett.* **114**, 217003 (2015).

32. Devereaux, T. P., Virosztek, A. & Zawadowski, A. Charge-transfer fluctuation, d-wave superconductivity, and the $B_{1g}$ Raman phonon in cuprates. *Phys. Rev. B* **51**, 505 (1995).

**Acknowledgments**

This work is supported by the U.S. Department of Energy (DOE), Office of Science, Basic Energy Sciences, Materials Sciences and Engineering Division, under contract DE-AC02-76SF00515. We acknowledge Diamond Light Source for providing the science commissioning time at the I21-RIXS beamline under Proposal SP18462.

**Author contributions:**

W. -S. L. conceived the experiment. W.-S.L., K.J.Z., M.H., J.L., A.N., A.C.W., M.G.-F. and H.R. conducted the experiment at Diamond Light Source. J.L. A.N. and K.J.Z performed the data processing from detector. W.-S.L. M.H, and H.L. analyzed the data. W.-S.L., B.M. J.Z. and T.P.D. performed the theoretical calculations. K.J.Z. A.C.W., and M.G.-F. constructed and commissioned the ID21-RIXS beamline and spectrometer at Diamond Light Source. M.Ha., D.S.

and H.E. synthesized and prepared samples for the experiments. W.-S.L., Z.X.S., B.M., J.Z., T.P.D. wrote the manuscript with input from all authors.

**Competing financial interests:**

The authors declare no competing financial interests.

**Data and materials availability:**

All experiment data is available in the main text or the supplementary materials. Codes and calculations used to support the plots or finding of this study is available from corresponding authors upon reasonable request.

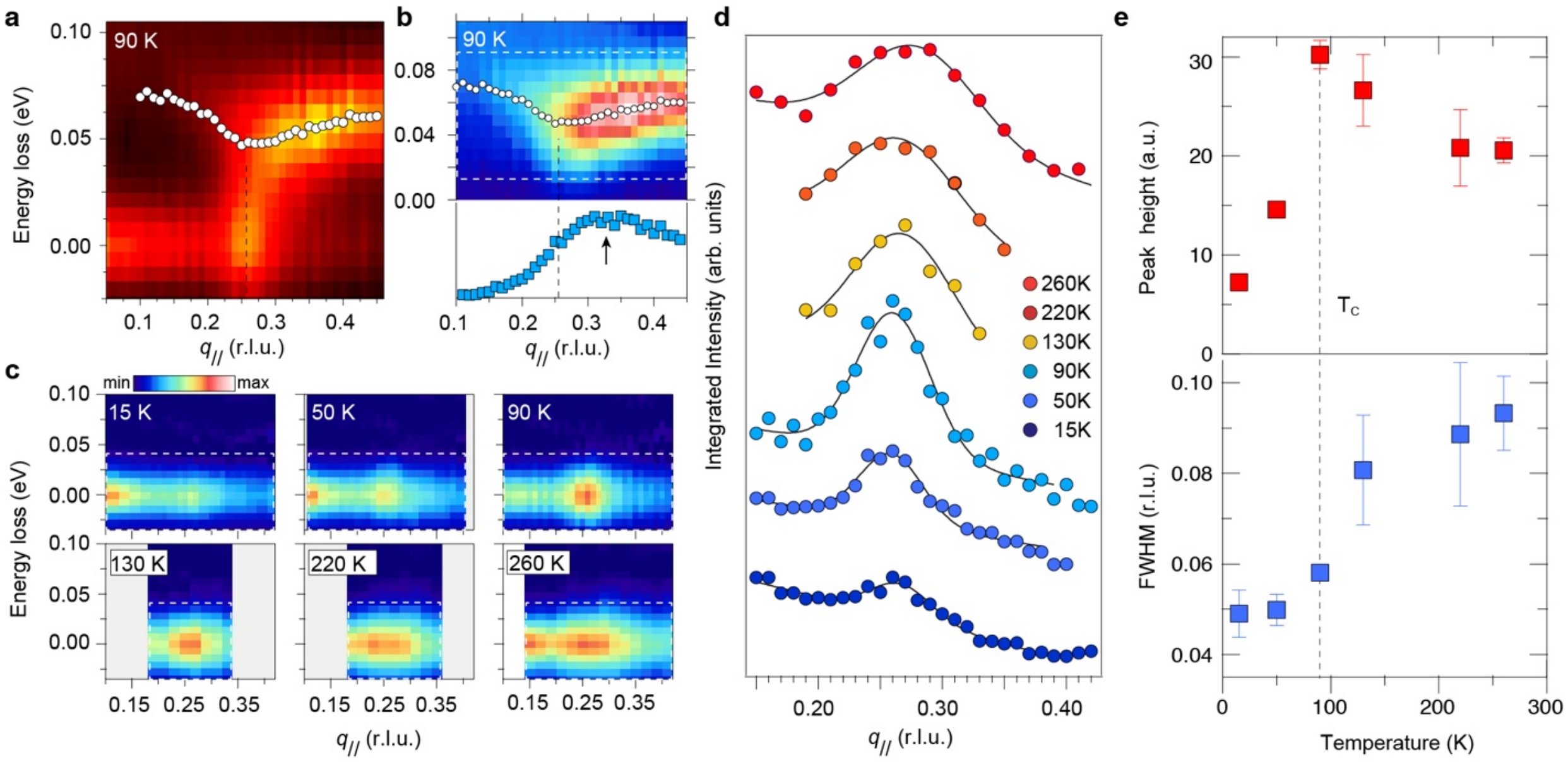

**Figure 1| Temperature dependence of the CO in the quasi-elastic region. a,** Raw RIXS intensity map along the (0,0)-(1,0) direction taken at $T_C$ (90K). White markers indicate the fitted peak positions of the excitations that possess an energy scale similar to bond-stretching phonons. The black dashed line indicates the CO wave-vector $Q_{CO}$. **b,** (Upper) RIXS phonon map obtained by subtracting the fitted elastic peak from raw data. (Lower) The momentum distribution curve of integrated phonon intensity within the white dashed box in the upper panel. The dashed line and arrow indicate $Q_{CO}$ and the maximum of the curve, respectively. **c,** Quasi-elastic maps taken at different temperatures obtained by subtracting the fitted phonon and background from raw data (Fig. 2a). **d**, CDW peak profiles obtained by integrating the quasi-elastic maps within the white dashed boxes in **c**. The black curves are fits to CDW peak profiles using a Gaussian function plus a linear background. **e,** Fitted intensity (upper) and the full width half maximum (FWHM, lower) of CDW peak profiles at different temperatures. The error bars were the standard deviation of the fit except for the 130K and 220K data, whose error bars were estimated via the uncertainty of the linear background.

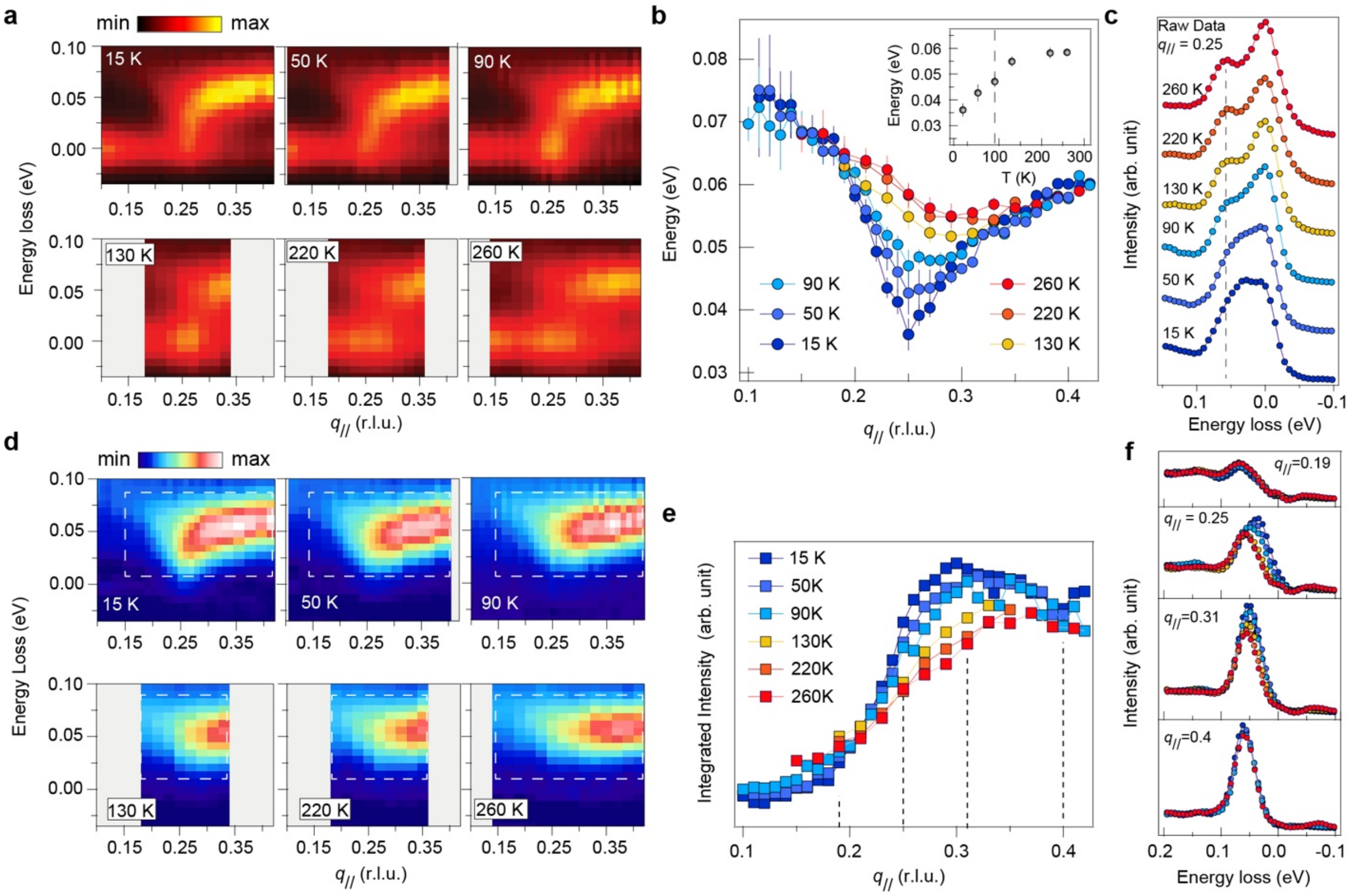

**Figure 2| The temperature dependence of RIXS phonon spectra. a,** Raw RIXS intensity maps taken at different temperatures. **b,** Fitted RIXS phonon dispersion at different temperatures. The inset shows the phonon energy at $q_{//}$ = 0.25 *r.l.u*. The dashed line indicates *Tc*. The error bars are estimated using 95% confidence interval of the fit. **c,** Raw energy loss spectrum at ~$Q_{CO}$ at different temperatures. The vertical dashed line indicates the phonon peak position at 260 K, serving as a guide-to-the-eye for the shift of position at low temperatures. **d,** RIXS phonon maps obtained by subtracting the fit of the elastic peaks from the raw data. **e**, Integrated momentum distribution curves at different temperatures, obtained by integrating RIXS phonon maps within the white dashed boxes indicated in (**d**). **f,** Temperature evolution of RIXS phonon spectra at representative $q_{//}$ as indicated by dashed lines in (**e**).

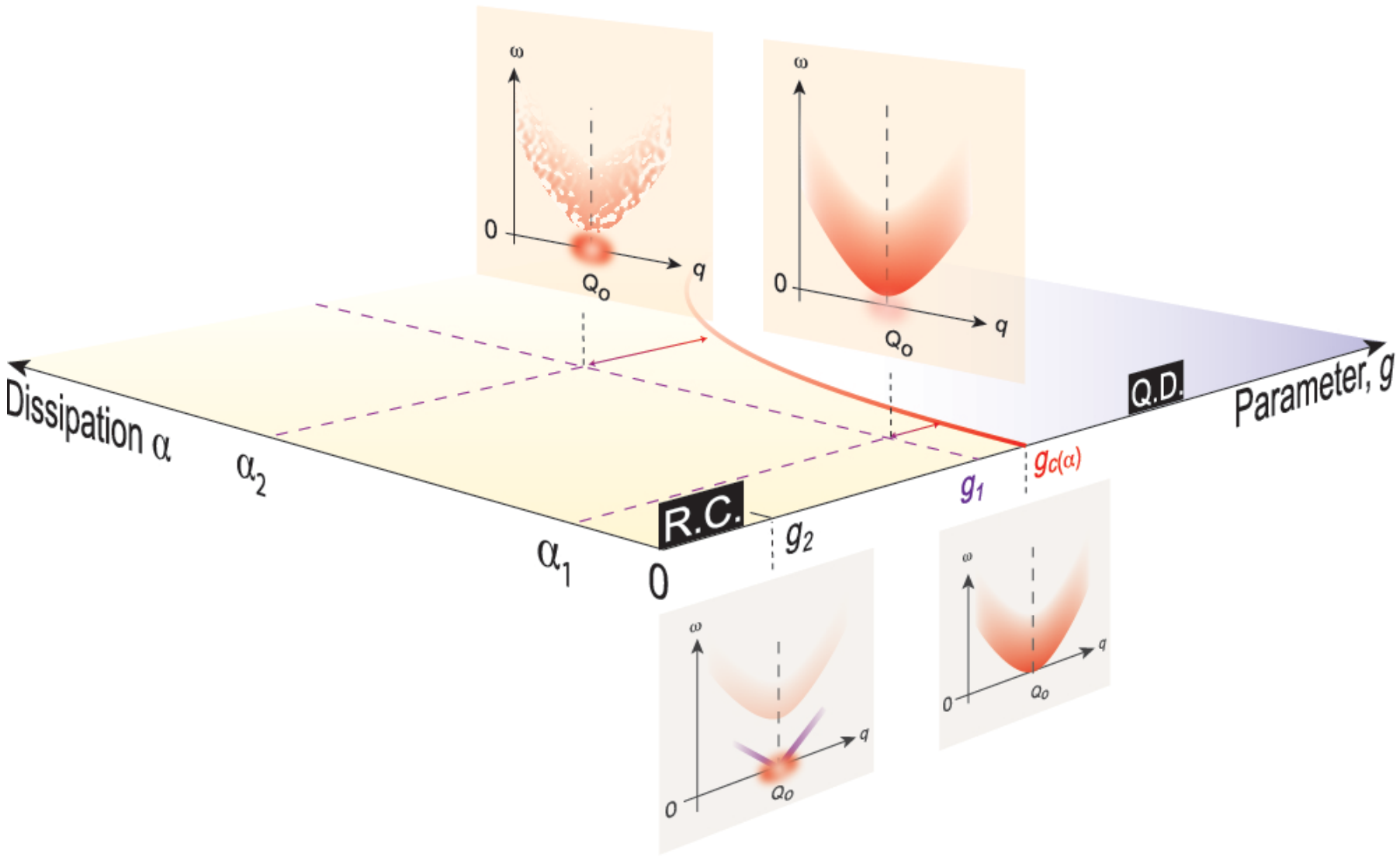

**Figure 3| Dissipation and Quantum Critical Point.** A sketch of the evolution of a CO order and the associated excitations (vertical panels) when approaching the quantum critical point ($g_c(\alpha)$) via changing a conceptual tuning parameter $g$ at zero dissipation (*e.g.* from $g_2$ to $g_C$) or varying "dissipation", (*e.g.* from $\alpha_1$ to $\alpha_2$) at $g_1$. The ordered side of the phase diagram is yellow shaded (renormalized classical, R.C.), while the disordered side is blue shaded (quantum disordered, Q.D.). The red round shades represent the static order at the energy $\omega = 0$ and the momentum $q = Q_0$, whose intensity represents the magnitude of the order parameter. The red shaded area in $\omega > 0$ represents the branch-cut continuum of quantum fluctuations associated with the order. The purple lines represent the Goldstone mode associated with the order. The graininess of the continuum at $\alpha_2$ illustrate its suppression from larger dissipation.

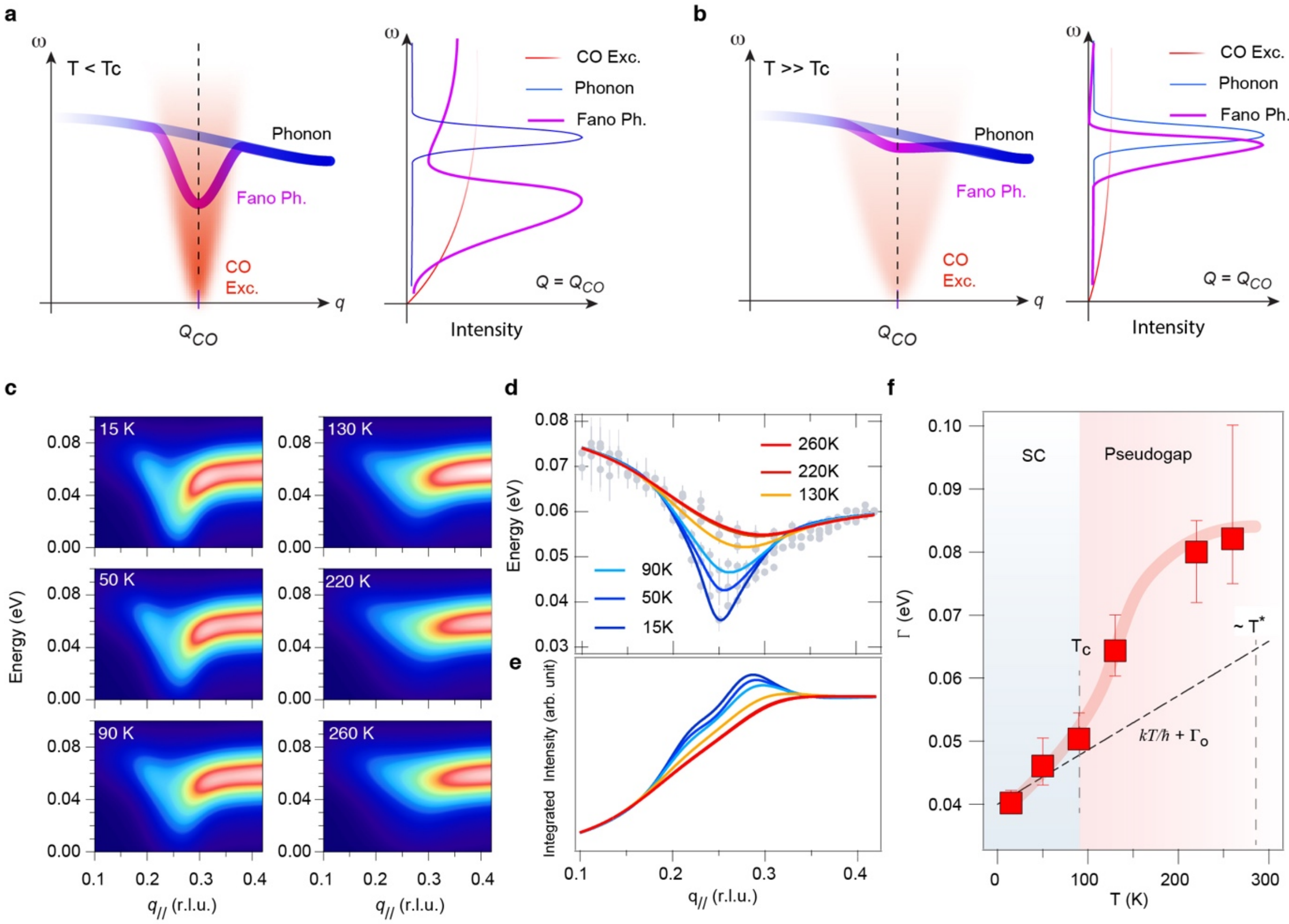

**Figure 4| Modeling the Fano interference between CO excitations and bond-stretching phonons. a,** (left) a sketch of the model to mimic the situation when T < $T_C$. The momentum-dependence of the electron-bond-stretching-phonon coupling strength were illustrated by the blue color gradient along the phonon dispersion curve. The CO excitation continuum (CO Exc.) were illustrated by the red shaded area. The dispersion modified by Fano-interference with CO excitation is sketched as the purple curve (Fano. Ph.). The right panel sketches the Fano spectrum at Q = $Q_{CO}$. **b,** a sketch of the model and Fano spectra for T >> $T_C$ where the CO is more damped in energy and broadened in momentum space, resulting a weaker Fano effect. **c,** Calculated phonon intensity maps for different temperatures. A Gaussian convolution of 35 meV FWHM is applied to account for the energy broadening due to instrument resolution in the experiment data. **d**, The peak positions of the Fano spectra shown in **C**. Grey markers are the experimental data of Fig. 2b. **e,** Integrated momentum distribution curves at different temperatures, obtained by integrating the maps shown in (c) from 0.01 to 0.09 eV. **f,** The damping parameter Γ at different temperatures. The vertical dashed lines indicate the superconducting transition ($T_c$) and pseudogap ($T^*$)

temperatures, respectively. The black dashed line indicates the Planckian scale ($kT/\hbar$) plus a zero-temperature residue $\Gamma_o$ = 0.04 eV. The inset shows the lifetime converted from the Γ.

# Supplementary Information for

## Spectroscopic Evidence for Charge Order Melting via Quantum Fluctuations in a Cuprate

W. S. Lee[1]*, K. J. Zhou[2]*, M. Hepting[1], J. Li[2], A. Nag[2], A. C. Walters[2], M. Garcia-Fernandez[2], H. Robarts[2], M. Hashimoto[3], H. Lu[4], B. Nosarzewski[4], D. Song[5], H. Eisaki[5], Z. X. Shen[1], B. Moritz[1], J. Zaanen[6], T. P. Devereaux[1]*

[1]Stanford Institute for Materials and Energy Sciences, SLAC National Accelerator Laboratory and Stanford University, 2575 Sand Hill Road, Menlo Park, California 94025, United States of America

[2]Diamond Light Source, Harwell Science and Innovation Campus, Didcot, Oxfordshire OX11 0DE, United Kingdom

[3]Stanford Synchrotron Radiation Lightsource, SLAC National Accelerator Laboratory, Menlo Park, CA 94025, United States of America

[4]Department of Physics, Stanford University, Stanford, California 94305, United States of America

[5]National Institute of Advanced Industrial Science and Technology (AIST), Tsukuba, Ibaraki 305-8560, Japan

[6]Geballe Laboratory for Advanced Materials, Departments of Physics and Applied Physics, Stanford University, Stanford, CA 94305, USA.

[7]Instituut-Lorentz for theoretical Physics, Leiden University, Niels Bohrweg 2, 2333 CA Leiden, The Netherlands.

Correspondence to: leews@stanford.edu, kejin.zhou@diamond.ac.uk, tpd@stanford.edu

This Supplementary Information contains:

Supplementary Figure 1 - 3

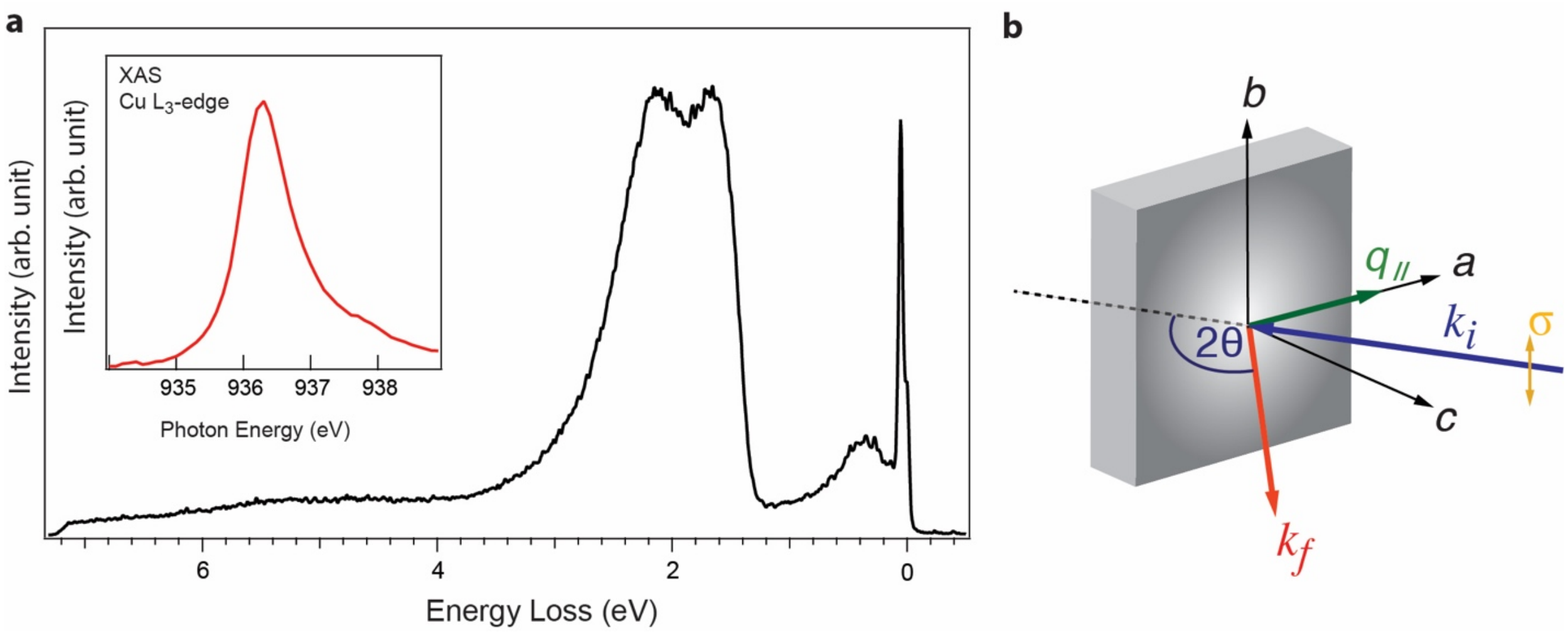

**Supplementary Figure 1| RIXS experiment and scattering geometry. a,** A representative RIXS spectral taking at Cu $L_3$-edge. In this work, we discuss the inelastic scattering signal below 0.1 eV. The inset shows a typical x-ray absorption curve across the Cu $L_3$-edge, taken by measuring total florescent yield. **b,** A sketch of the scattering geometry. a, b, and c represent the crystal axes. $k_i$ and $k_f$ represent the incident and scattering photon momentum. 2θ is the scattering angle. The yellow arrow represents the polarization of the incident x-ray. All the data shown in this manuscript were taken using incident beam with sigma polarization.

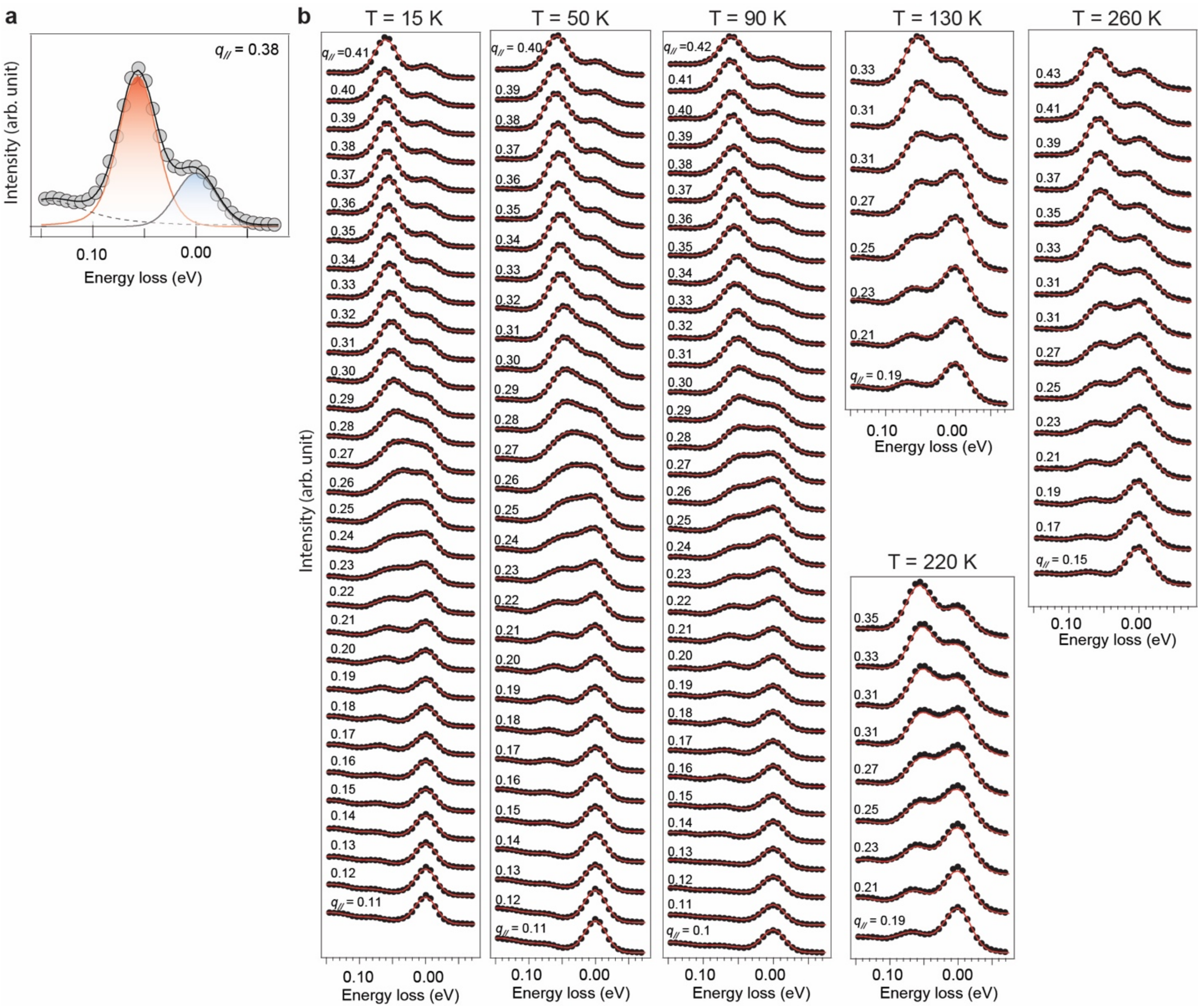

**Supplementary Figure 2| Raw RIXS spectra and Fitting. a,** An example of the spectra fitting. The fit function consists of a Gaussian function for the elastic peak (blue), an anti-Lorentzian function for the RIXS phonon (red), and a smooth background from high energy (black dashed line). The fit function is convoluted with a Gaussian function with a FWHM corresponding to the energy resolution of the RIXS instrument and fit to the data. **b,** All Raw RIXS spectra discussed in this work. The red curves are the fits, showing good agreements with the experimental data.

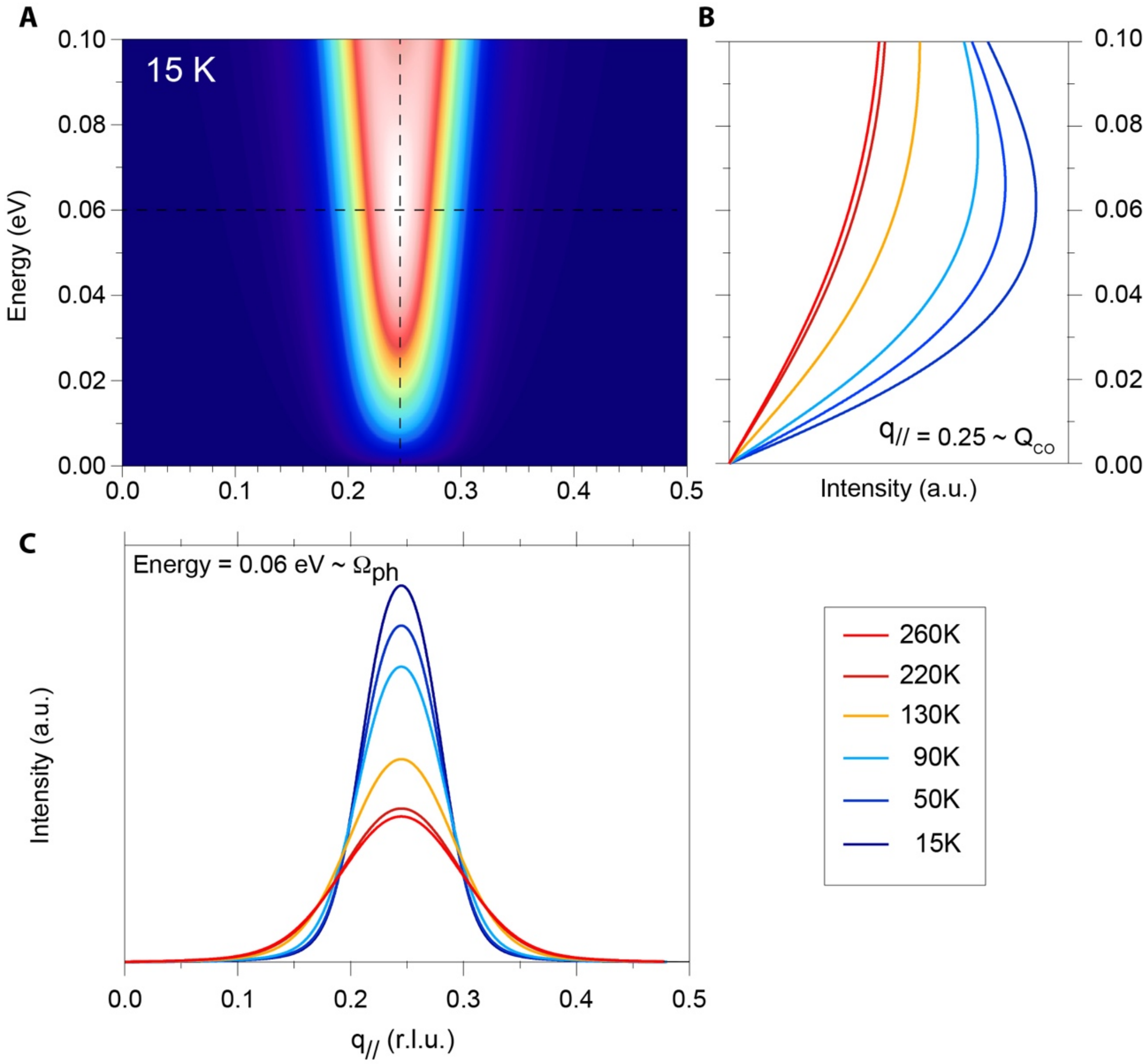

**Supplementary Figure 3| CO excitations continuum used in the model. a,** Intensity map of $Im[\chi(q,\omega)]$ with for modeling 15 K data. **b,** Energy distribution curves at $Q_{CO}$ (the vertical black dashed lines in **a**) with different values of the damping Γ that are used to model the temperature dependence behavior shown in Fig. 4c**. c,** Momentum distribution curve at 0.06 eV (the horizontal black dashed lines in **a**), approximately at the bond-stretching phonon energy.